\documentclass[11pt,a4paper]{article}

\usepackage[mathcal]{eucal}
\usepackage{color}
\usepackage{latexsym}
\usepackage{amsfonts}
\usepackage{amsmath}
\usepackage{amsthm}
\usepackage{amssymb}
\usepackage{mathrsfs}
\usepackage{graphicx}
\usepackage{algorithm}
\usepackage{algorithmic}

\newtheorem{theorem}{Theorem}[section]
\newtheorem{lemma}[theorem]{Lemma}

\newenvironment{program}
	{\small  \begin{eqnarray*} } { \end{eqnarray*} }

\newlength{\programindent}
\newcommand{\prstyle}[1]{\mbox{\bf #1}}
\newcommand{\ind}[1]	{\hspace*{#1\programindent}}
\newcommand{\prline}[2]	{&& \ind{#1} #2 \\}

\newcommand{\Assign}[2]
	{#1 \leftarrow #2;}
\newcommand{\AssignNoDelim}[2]
	{#1 \leftarrow #2}
\newcommand{\Foralldo}[1]
	{\prstyle{for~all~} #1 \prstyle{~do~}}

\newcommand{\Whiledo}[1]
	{\prstyle{while~} #1 \prstyle{~do~}}

\newcommand{\Function}[2]
	{ \mbox{\rm #1}(#2)}
\newcommand{\Return}[1]
	{\prstyle{return~} #1;}

\newcommand{\BIGOP}[1]{\mathop{\mathchoice%
{\raise-0.22em\hbox{\huge $#1$}}%
{\raise-0.05em\hbox{\Large $#1$}}{\hbox{\large $#1$}}{#1}}}

\DeclareMathOperator*{\pr}{Pr}

\begin{document}

\title{A constructive proof of the general Lov\'asz Local Lemma}
\author{Robin A. Moser\thanks{Research is
   supported by the SNF Grant 200021-118001/1}
\\ \vspace{0.1cm} \\
\small
Institute for Theoretical Computer Science\\
\small
Department of Computer Science\\
\small
ETH Z\"urich, 8092 Z\"urich, Switzerland\\
\small
\texttt{robin.moser@inf.ethz.ch}
\and
G\'abor Tardos\thanks{Supported by NSERC grant 329527, and by OTKA grants
T-046234, AT-048826 and NK-62321}
\\ \vspace{0.1cm} \\
\small
School of Computing Science\\
\small
Simon Fraser University\\
\small
Burnaby, BC, Canada\\
\small
and\\
\small
R\'enyi Institute\\
\small
Budapest, Hungary\\
\small
\texttt{tardos@cs.sfu.ca}
}
\date{May 2009}
\maketitle

\begin{abstract} 
The Lov\'asz Local Lemma \cite{EL75} is a powerful tool to non-constructively prove the existence of combinatorial objects meeting
a prescribed collection of criteria. In his breakthrough paper \cite{Bec91}, Beck demonstrated that a constructive variant can be
given under certain more restrictive conditions. Simplifications of his procedure and relaxations of its restrictions were subsequently 
exhibited in several publications \cite{Alo91,MR98,CS00,Mos06,Sri08,Mos08a}. In \cite{Mos08b}, a constructive proof was presented that works under
negligible restrictions, formulated in terms of the Bounded Occurrence Satisfiability problem. In the present paper, we reformulate
and improve upon these findings so as to directly apply to almost all known applications of the general Local Lemma.
\end{abstract}


\section{Introduction}

Let $\mathcal A$ be a finite collection of mutually independent events in a
probability space. The probability that none of these events happen is exactly
$\prod_{A\in\mathcal A}(1-\pr[A])$. In particular, this probability is
positive whenever no event in $\mathcal A$ has probability $1$. L\'aszl\'o
Lov\'asz's famous Local Lemma \cite{EL75} allows for limited dependence
among the events, but still concludes that with positive probability
none of the events happen if the individual events have bounded
probability. Here is the lemma in a very general form.

\begin{theorem}\label{lovasz}
{\bf \cite{EL75}}
Let $\mathcal A$ be a finite set of events in a probability space. For
$A\in\mathcal A$ let $\Gamma(A)$ be a subset of $\mathcal A$ satisfying that
$A$ is independent from the collection of events $\mathcal
A\setminus(\{A\}\cup\Gamma(A))$. If there exists an
assignment of reals $x : \mathcal{A} \rightarrow (0,1)$ such that
$$ \forall A \in \mathcal{A} \; : \; \pr[A] \; \le \; x(A) \prod_{B \in
\Gamma(A)} (1 - x(B)),$$
then the probability of avoiding all events in $\mathcal A$ is at least
$\prod_{A\in\mathcal A}(1-x(A))$, in particular it is positive.
\end{theorem}

The original proof of this statement is non-constructive and does not yield an efficient procedure for
searching the probability space for a point with the desired property. The purpose of the present paper is to give 
an alternative, algorithmic proof that provides such a procedure. This is not the first attempt to do
so. In \cite{Bec91}, Beck achieved a significant breakthrough demonstrating that algorithmic versions of the
Local Lemma exist. He formulated his strategy in terms of hypergraph 2-coloring as a specific application
of the lemma and proved that if in a hypergraph, every edge contains at least $k$ vertices and shares common
vertices with no more than roughly $2^{k/48}$ other edges, then a polynomial time algorithm can 2-color the
vertices without producing a monochromatic edge. The existential version of the Local Lemma on the other hand
allows for every edge to share vertices with roughly $2^{k}/e$ other edges and guarantees the existence of such
a coloring. Subsequently, several authors have attempted to improve upon the gap between the existential version
and its constructive counterparts. Alon improved the threshold to essentially $2^{k/8}$ 
using a simpler and randomized variant of Beck's algorithm \cite{Alo91}. Molloy and Reed provided in \cite{MR98}
a general framework capturing the requirements a particular application has to meet so as to become tractable by
the tools of Beck and Alon. A small error in this result was recently fixed in
\cite{PT09}. Czumaj and Scheideler's contribution in \cite{CS00} extended known algorithmic versions
to somewhat more general cases where the edges of the hypergraph in question need not be of uniform size. Srinivasan 
in \cite{Sri08} provided another improvement that reduced the gap to a threshold of essentially $2^{k/4}$ along with
a series of other advantages over the previous approaches. In \cite{Mos08a}, yet another variant was presented
that achieves a polynomial running time for instances up to a neighborhood size of roughly $2^{k/2}$ and finally
in \cite{Mos08b}, the threshold was lowered to roughly $2^{k}/32$. In the present paper,
we reformulate and improve upon the last cited result both so as to get rid of the now unnecessary constant in the
hypothesis and so as to directly apply to almost all applications of the Local Lemma known so far. The only restriction
we have to impose upon the general setting in the non-constructive version as formulated above will be that we
consider events determined by different subsets of underlying mutually independent random variables
and $\Gamma(A)$ consists of all events that depend on some of the same
variables as $A$. See the exact formulation below. While this appears to be necessary in order to get any algorithmic access to the
problem, it seems as well to be the case in almost all known applications. 

Let $\mathcal P$ be a finite collection of mutually independent random
variables in a fixed probability space $\Omega$. We will consider events $A$
that are determined by the values of some subset $S\subseteq\mathcal P$ of
these variables. In such a case we say that an evaluation of the variables
in $S$ \emph{violates} $A$ if it makes $A$ happen. Clearly, if $A$ is
determined by $\mathcal P$, then there is a unique minimal subset
$S\subseteq\mathcal{P}$ that determines $A$. We denote this set of variables by
$\mbox{vbl}(A)$ and assume throughout the paper that this set is given to
all algorithms dealing with the event $A$. 

Let $\mathcal{A}$ be a finite family of events in
$\Omega$ determined by $\mathcal P$. We define the \emph{dependency graph}
$G=G_{\mathcal A}$ for $\mathcal A$ to be the graph on vertex set $\mathcal{A}$ with an edge
between events $A,B \in \mathcal{A}$ if
$A \ne B$ but $\mbox{vbl}(A) \cap \mbox{vbl}(B) \ne \emptyset$.
For $A\in \mathcal A$ we write $\Gamma(A)=\Gamma_{\mathcal A}(A)$ for the neighborhood of $A$ in $G$. Note that $\Gamma(A)$ satisfies the requirement in
Theorem~\ref{lovasz} as $A$ is determined by the variables in $\mbox{vbl}(A)$
and the events in $\mathcal A\setminus(\{A\}\cup\Gamma(A))$ are determined by
the rest of the variables in $\mathcal P$.

Given the family $\mathcal A$ of events as above our goal is not only to show
that there exists an evaluation that does not violate any event in the family
but to efficiently find such an evaluation. The algorithm we suggest (Algorithm 1.1) 
is as simple and natural as it can get: We start with a random point in $\Omega$ and
maintain an evaluation $v_P$ of each variable $P \in \mathcal P$. We check whether
some event in $\mathcal A$ is violated. If so, we arbitrarily pick a violated
event $A\in\mathcal A$ and sample another random assignment of values for the
variables in $\mbox{vbl}(A)$ on which $A$ depends, each one independently and
according to its distribution while not changing the values of the variables in
$\mathcal P\setminus\mbox{vbl}(A)$. We call this a \emph{resampling} of the
event $A$. We continue resampling violated events until no such event exists
anymore. We will prove that this simple algorithm quickly terminates, i.e., it
quickly reaches an evaluation of the variables not violating any of the events in
$\mathcal A$ if the conditions of the Local Lemma are satisfied.


\begin{center}
\begin{minipage}{0.75\textwidth}
\hrule
\vspace*{-0.2cm}
\begin{program}
\prline{0}{\prstyle{function~} \Function{sequential\_lll}{\mathcal P, \mathcal A}}
\prline{1}{\Foralldo{P \in \mathcal{P}}}
\prline{2}{\Assign{v_P}{\mbox{a random evaluation of } P}}
\prline{1}{\Whiledo{\exists A \in \mathcal A : A \mbox{ is violated when }
    (P=v_P : \forall P \in \mathcal P) }}
\prline{2}{\mbox{pick an arbitrary violated event }A\in\mathcal A;}
\prline{2}{\Foralldo{P \in \mbox{vbl}(A)}}
\prline{3}{\Assign{v_P}{\mbox{a new random evaluation of } P }}
\prline{1}{\Return{(v_P)_{P \in \mathcal P}}}
\end{program}
\vspace*{-1cm}
\hrule
\center{
\emph{Algorithm 1.1: the sequential solver}
}
\end{minipage}
\end{center}


The efficiency of the method clearly depends upon whether random values for each variable can be efficiently sampled and
whether they can be efficiently checked against the given events. This is the case for almost all known applications of
the lemma and it is less restrictive than previously known methods which required
conditional probabilities or expectations to be computed. We will analyze the
efficiency of the algorithm in terms of the expected
number of times an event $A\in\mathcal A$ is resampled.

\begin{theorem} 
\label{thmmain}
Let $\mathcal P$ be a finite set of mutually independent random variables in a
probability space. Let $\mathcal{A}$ be a finite set of events determined by
these variables. If there exists an assignment of reals $x : \mathcal{A} \rightarrow (0,1)$ such that
$$ \forall A \in \mathcal{A} \; : \; \pr[A] \; \le \; x(A)
\hspace*{-0.2cm} \prod_{B \in \Gamma_{\mathcal A}(A)} (1 - x(B)),$$
then there exists an assignment of values to the variables $\mathcal P$
not violating any of the events in $\mathcal A$. Moreover the
randomized algorithm described above resamples an event $A \in \mathcal{A}$ at
most an expected $x(A)/(1-x(A))$ times before it finds such an
evaluation. Thus the expected total number of resampling steps is at most
$\sum_{A \in \mathcal{A}} \frac{x(A)}{1-x(A)}.$ 
\end{theorem}

Our algorithm lends itself for parallelization. In the \emph{parallel version}
of the algorithm (Algorithm 1.2) we start again with the evaluation of the variables 
at a random point in the probability space, then in every step we select a maximal
independent set $S$ in the subgraph of the dependency graph $G$ spanned by the
violated events and resample all the variables these events depend on in
parallel. That is, we take independent new samples of the variables in
$\cup_{A\in S}\mbox{vbl}(A)$ and keep the values assigned to the rest of the
variables. We continue until we find an evaluation not violating any of the
events. This algorithm can be considered a
special case of the sequential algorithm, so the statement of
Theorem~\ref{thmmain} applies to the parallel version too. In order to give a
logarithmic bound for the expected number of steps we assume slightly stronger
bounds on the probabilities of the events.


\begin{center}
\begin{minipage}{0.75\textwidth}
\hrule
\vspace*{-0.2cm}
\begin{program}
\prline{0}{\prstyle{function~} \Function{parallel\_lll}{\mathcal P, \mathcal A}}
\prline{1}{\Foralldo{P \in \mathcal{P}} \prstyle{in parallel}}
\prline{2}{\Assign{v_P}{\mbox{a random evaluation of } P}}
\prline{1}{\Whiledo{\exists A \in \mathcal A : A \mbox{ is violated when }
    (P=v_P : \forall P \in \mathcal P) }}
\prline{2}{\AssignNoDelim{S}{\mbox{a maximal independent set in the subgraph of $G_{\mathcal{A}}$ induced by all events}}}
\prline{2}{\hspace*{0.8cm} \mbox{which are violated when } (P=v_P : \forall P \in \mathcal P), \mbox{ constructed in parallel};}
\prline{2}{\Foralldo{P \in \bigcup_{A \in S} \mbox{vbl}(A)} \prstyle{in parallel}}
\prline{3}{\Assign{v_P}{\mbox{a new random evaluation of } P }}
\prline{1}{\Return{(v_P)_{P \in \mathcal P}}}
\end{program}
\vspace*{-1cm}
\hrule
\center{
\emph{Algorithm 1.2: the parallel solver}
}
\end{minipage}
\end{center}


\begin{theorem} 
\label{parallel}
Let $\mathcal P$ be a finite set of mutually independent random variables in a
probability space. Let $\mathcal{A}$ be a finite set of events determined by
these variables. If $\varepsilon>0$ and there exists an assignment of reals $x : \mathcal{A} \rightarrow (0,1)$ such that
$$ \forall A \in \mathcal{A} \; : \; \pr[A] \; \le \; (1-\varepsilon) x(A)
\hspace*{-0.2cm} \prod_{B \in \Gamma_{\mathcal A}(A)} (1 - x(B)),$$
then the parallel version of our algorithm takes an expected
$O(\frac1\varepsilon\log\sum_{A\in\mathcal A}\frac{x(A)}{1-x(A)})$ steps before it finds an evaluation violating no event in $\mathcal A$.
\end{theorem}

There is not much about our algorithm that is inherently randomized. We can demonstrate
that under additional conditions, most notably a constant bound on the maximum degree of
the dependency graph, the same task can be performed by a deterministic procedure.

\begin{theorem}
\label{deterministic}
Let $\mathcal P = \{P_1, P_2, \mathellipsis, P_n\}$ be a finite set of
mutually independent random variables in a probability space, each
$P_i$ taking values from a finite domain $D_i$. Let
$\mathcal{A}$
be a set of $m$ events determined by these variables. Consider the
problem size to be $s := m+n+\sum_{i =1}^n |D_i|$. Suppose there
exists an algorithm that can compute, for each $A \in \mathcal{A}$ and
each partial evaluation $(v_i \in D_i)_{i \in I}$, $I \subseteq [n]$
the conditional probability $\mbox{Pr}[A | \forall i \in I : P_i=v_i]$
in time polynomial in $s$. Suppose, moreover, that the maximum degree of the dependency graph
$G_{\mathcal{A}}$ is bounded by a constant,
that is $\forall A \in \mathcal{A} : |\Gamma_{\mathcal{A}}(A)| \le k$ for some
constant $k$. If there is a constant $\varepsilon > 0$ and
an assignment of reals $x : \mathcal{A} \rightarrow (0,1)$ such that 
$$ \forall A \in \mathcal{A} \; : \; \pr[A] \; \le \; (1-\varepsilon) x(A)
\hspace*{-0.2cm} \prod_{B \in \Gamma_{\mathcal A}(A)} (1 - x(B)),$$
then a deterministic algorithm can find an evaluation of the variables
such that no event occurs in time polynomial in $s$.
\end{theorem}

In Sections~2 and 3 we provide an analysis of the algorithm, proving
Theorem~\ref{thmmain}, in Section~4 we analyze the parallel version and prove
Theorem~\ref{parallel}. In Section~5,
we prove the correctness of the derandomized algorithm as claimed in
Theorem~\ref{deterministic}. Section~6 contains a lopsided version of
Theorem~\ref{thmmain} and Section~7 has concluding remarks. 

\section{Execution logs and witness trees}
\label{secalgo}

Note that the decision which violated event $A \in \mathcal{A}$ to correct
in each step can be taken completely arbitrarily. Let us fix any
(deterministic or randomized) procedure for this selection, this makes the
algorithm and the expected values we consider well defined.
The selection method the algorithm uses does not matter for our
analysis.

We need to record an accurate journal of what the algorithm does. Let $C : \mathbb{N} \rightarrow
\mathcal{A}$ list the events as they have been selected for resampling in each step. If the algorithm terminates, $C$ is partial and defined only up
to the given total number of steps carried out. We call $C$ the \emph{log} of the execution. With a fixed selection discipline as
described above, $C$ is now a random variable determined by the random choices
the algorithm makes.

Recall that dependency graph $G$ is a graph on vertex set $\mathcal A$ where
two distinct events $A,A'\in\mathcal A$ are connected if
$\mbox{vbl}(A)\cap\mbox{vbl}(A')\ne\emptyset$ and that $\Gamma(A)$ denotes the
neighborhood of the vertex $A$ in $G$. We also use here the
\emph{inclusive neighborhood} $\Gamma^+(A) := \Gamma(A) \cup \{A\}$ of a
vertex $A$.

A \emph{witness tree} $\tau = (T, \sigma_T)$ is a finite rooted tree $T$ together with a labelling $\sigma_T : V(T) \rightarrow \mathcal{A}$ 
of its vertices with events such that the children of a vertex $u \in V(T)$
receive labels from $\Gamma^+(\sigma_T(u))$. If distinct children of the same
vertex always receive distinct labels we call the witness tree \emph{proper}.
To shorten notation, we will write $V(\tau) := V(T)$ and for any $v \in V(\tau)$, we write $[v] := \sigma_T(v)$.
Given the log $C$, we will now associate with each resampling step $t$ carried out a witness tree $\tau_C(t)$ 
that can serve as a `justification' for the necessity of that correction
step. Let us define $\tau_C^{(t)}(t)$ to be an isolated root
vertex labelled $C(t)$. Then going backwards through the log, for each $i=
t-1, t-2, \mathellipsis, 1$, we distinguish two cases. If there
is a vertex $v \in \tau_C^{(i+1)}(t)$ such that $C(i) \in \Gamma^+([v])$,
then we choose among all such vertices the one having the maximum distance
from the root and attach a new child vertex $u$ to $v$ that we label $C(i)$, thereby obtaining
the tree $\tau_C^{(i)}(t)$. In the selection of the maximum distance vertex we
break ties arbitrarily. If there is no vertex $v \in \tau_C^{(i+1)}(t)$ such
that $C(i) \in \Gamma^+([v])$, then we skip time step $i$ and simply define $\tau_C^{(i)}(t) := \tau_C^{(i+1)}(t)$.
Finally let $\tau_C(t) := \tau_C^{(1)}(t)$.

We say that the witness tree $\tau$ \emph{occurs} in the log $C$
if there exists $t \in \mathbb{N}$ such that $\tau_C(t) = \tau$.

\begin{lemma}
\label{coupling}
\renewcommand{\labelenumi}{\roman{enumi}.}
Let $\tau$ be a fixed witness tree and $C$ the (random) log produced by the algorithm.
\begin{description}
 \vspace*{-0.2cm}
 \item[(i)] If $\tau$ occurs in $C$, then $\tau$ is proper.
 \vspace*{-0.2cm}
 \item[(ii)] The probability that $\tau$ appears in $C$ is at most $\prod_{v\in V(\tau)}\pr[[v]]$.
\end{description}
\end{lemma}

\emph{Proof.}
Assume $\tau$ occurs in the log $C$, so we have $\tau_C(t)=\tau$ for some
$t\in \mathbb N$. For a vertex $v\in V(\tau)$ let $d(v)$ denote the
\emph{depth} of vertex $v$, that is its distance from the root and let $q(v)$ stand for the step of
the algorithm constructing $\tau_C(t)$ in which $v$ was attached, that is,
$q(v)$ is the largest value $q$ with $v$ contained in $\tau_C^{(q)}(t)$.

First note that if $q(u)<q(v)$ for vertices $u,v\in V(\tau)$ and
$\mbox{vbl}([u])$ and $\mbox{vbl}([v])$ are not disjoint, then $d(u)>d(v)$. Indeed, when adding the vertex $u$
to $\tau_C^{(q(u)+1)}(t)$ we attach it to $v$ or to another vertex of equal or
greater depth. As a consequence observe that for any two vertices $u, v \in V(\tau)$ 
at the same depth $d(v)=d(u)$, $[u]$ and $[v]$ do not depend on any common variables, that is the labels in every level of $\tau$
form an independent set in $G$. In particular $\tau$ must be proper, establishing claim (i).

Consider the following procedure that we call $\tau$-check:
In an order of decreasing depth (e.g., reversed breadth first search order)
visit the vertices of $\tau$ and for a
vertex $v$ take a random evaluation of the variables in $\mbox{vbl}([v])$
(according to their distribution, independent of possible earlier evaluations)
and check if the resulting evaluation violates $[v]$. We say that the
$\tau$-check passes if all events were violated when checked.

Trivially, the $\tau$-check passes with probability exactly $\prod_{v\in
V(\tau)}\pr[[v]]$. The lemma follows from the observation that whenever
$\tau$ occurs in the log and we run the $\tau$-check on the same random source it
passes. For this coupling argument we have to specify explicitly
how the algorithms use the random source. We assume that for
each variable $P\in\mathcal P$ the random source produces an infinite list of
independent random samples $P^{(0)},P^{(1)},\ldots$, and whenever either algorithm
calls for a new random sample of $P$ we pick the next unused
value from this sequence.

Having assumed that $\tau=\tau_C(t)$, we need to prove that the $\tau$-check
passes, that is, when it considers a vertex $v\in V(\tau)$ and takes random
samples of the variables in $\mbox{vbl}([v])$, the resulting evaluation
violates $[v]$. Let us fix the vertex $v$ and for $P\in\mbox{vbl}([v])$ let
$S(P)$ be the set of vertices $w\in V(\tau)$ with $d(w)>d(v)$ and $P\in \mbox{vbl}([w])$. When the
$\tau$-check considers the vertex $v$ and samples $P$ the random source
gives $P^{(|S(P)|)}$. This is because the $\tau$-check visits the vertices in order of
decreasing depth and among the vertices with depth equal to $d(v)$ only the
label of $v$ depends on $P$, so before the $\tau$-check considers $v$ it had
sampled $P$ exactly when it was considering the vertices in $S(P)$.

In step $q(v)$, our algorithm chooses the event
$[v]$ for resampling, so $[v]$ must be violated before this resampling. We claim that for $P\in\mbox{vbl}([v])$ the
current value of the variable $P$ is $P^{(|S(P)|)}$ at this time. Indeed,
$P$ was sampled at the beginning of the algorithm and then at the steps
$q(w)<q(v)$ for $w\in S(P)$. As the $\tau$-check has these exact same values
for the variables in $\mbox{vbl}([v])$ when considering $v$ it also must find
that $[v]$ is violated, proving (ii). \qed

For any event $A \in \mathcal{A}$, let us denote by $N_A$ the random variable
that counts how many times the event $A$ is resampled during the execution of
our algorithm. If $C$ is the log of the execution of our algorithm, then $N_A$
is the number of occurrences of $A$ in this log and also the number of distinct
proper witness trees occurring in $C$ that have their root labeled $A$. The
latter statement holds because if $t_i$ is the $i$-th time step
with $C(t_i)=A$, then obviously the tree $\tau_C(t_i)$ contains exactly $i$
vertices labelled $A$, thus $\tau_C(t_i)\ne\tau_C(t_j)$ unless
$i=j$. Therefore one can bound the expectation of $N_A$ simply by summing the
bounds in Lemma~\ref{coupling} on the probabilities of the occurrences of the
different proper witness trees. In the next section we do just that by
relating these probabilities to a random process.

\section{Random generation of witness trees}

Let us fix an event $A \in \mathcal{A}$ and consider the following multitype
Galton-Watson branching process for generating a proper witness tree having its root labelled $A$. 
In the first round, we produce a singleton vertex labelled $A$. Then in each subsequent round, we consider each vertex $v$ produced in the
previous round independently and, again independently, for each event $B \in \Gamma^+([v])$ identical or adjacent to $[v]$ in the dependency graph, 
we add to $v$ a child node carrying the label $B$ with probability $x(B)$ or
skip that label with probability $1-x(B)$. All these choices are independent. The process continues until
it dies out naturally because no new vertices are born in some round (depending on the probabilities used, there is, of course, the possibility that
this never happens).

Let $x'(B):=x(B)\prod_{C\in\Gamma(B)}(1-x(C))$. For the probability that the
described Galton-Watson process yields a prescribed proper witness tree
we obtain the following.

\begin{lemma}
\label{lemgwprob}
Let $\tau$ a fixed proper witness tree with its root vertex labelled $A$. The
probability $p_\tau$ that the Galton-Watson process described above yields
exactly the tree $\tau$ is
$$ p_\tau =  \frac{1-x(A)}{x(A)} \prod_{v \in V(\tau)}x'([v]).$$
\end{lemma}

\emph{Proof.} For a vertex $v \in V(\tau)$ we denote by $W_v \subseteq \Gamma^+([v])$ the set of inclusive neighbors
of $[v]$ that do not occur as a label of some child node of $v$. Then clearly, the probability that the Galton-Watson process produces exactly
$\tau$ is given by
$$ p_\tau \; = \; \frac{1}{x(A)} \prod_{v \in V(\tau)} \left( x([v]) \prod_{u \in W_v} (1-x([u])) \right), $$
where the leading factor accounts for the fact the the root is always born. In order to get rid of the $W_v$, we can rewrite this 
expression in an obvious way to obtain
$$ p_\tau \; = \; \frac{1-x(A)}{x(A)} \prod_{v \in V(\tau)} \left( \frac{x([v])}{1-x([v])} \prod_{u \in \Gamma^+([v])} (1-x([u])) \right), $$
where again we have to account for the root separately. Replacing inclusive by exclusive neighborhoods, this simplifies to
$$ p_\tau \; = \; \frac{1-x(A)}{x(A)} \prod_{v \in V(\tau)} \left( x([v]) \prod_{u \in \Gamma([v])} (1-x([u])) \right)=  \frac{1-x(A)}{x(A)} \prod_{v \in V(\tau)}x'([v]).$$
\qed

Let $\mathcal{T}_A$ denote the set of all proper witness trees having the root
labelled $A$. We have
$$ \mathbb{E}(N_A) = \sum_{\tau \in \mathcal{T}_A} \pr[\tau \mbox{
  appears in the log } C] \; \le \; \sum_{\tau \in \mathcal{T}_A}
  \prod_{v \in V(\tau)} \pr[[v]] \; \le \; \sum_{\tau\in
\mathcal{T}_A}\prod_{v\in V(\tau)}x'([v]),$$
where the first inequality follows from Lemma~\ref{coupling}, while the second
follows from the assumption in Theorem~\ref{thmmain}. We further have
$$\mathbb{E}(N_A)\;\le\;\sum_{\tau\in \mathcal{T}_A}\prod_{v\in
      V(\tau)}x'([v])\;=\;\frac{x(A)}{1-x(A)}\sum_{\tau\in\mathcal{T}_A}p_\tau\;\le\;\frac{x(A)}{1-x(A)},$$
where the equality comes from Lemma~\ref{lemgwprob}, while the last inequality
follows from the fact that the Galton-Watson process produces exactly one tree
at a time (not necessarily one from $\mathcal{T}_A$ since it might also grow
infinite). This concludes the proof of Theorem \ref{thmmain}. \qed

\section{Analyzing the parallel algorithm}

Let us consider an arbitrary execution of the parallel version of the
algorithm. We choose an arbitrary ordering of the violated events selected
for resampling at each step and consider that these resamplings are done in that
order sequentially. This way
we obtain an execution of the sequential algorithm. Let $S_j$ be the segment
of the log $C$ of this execution that corresponds to resamplings done in
step $j$ of the parallel algorithm. We call the maximal depth of a vertex in a
witness tree the \emph{depth} of the tree.

\begin{lemma}
\label{para}
If $t\in S_j$, then the depth of $\tau_C(t)$ is $j-1$.
\end{lemma}

\emph{Proof.} Let $t_k$ be the first number in the segment $S_k$ and let
$\tau_k=\tau_C^{(t_k)}(t)$ for $k\le j$. As the events resampled in the $j$-th parallel step are
independent, the root is the only vertex of $\tau_j$. For $k<j$ we obtain
$\tau_k$ from $\tau_{k+1}$ by attaching some vertices
corresponding to the $k$-th parallel step of the algorithm. As these vertices
have independent labels they can only add one to the depth. To see that they
do add one to the depth consider a vertex $v$ of $\tau_{k+1}$ of maximal
depth. This vertex
corresponds to a resampling of the event $[v]$ some time after step $k$ of the
parallel algorithm. If $\tau_k$ has no vertex with higher depth than $v$, then
from the parallel step $k$ to the resampling corresponding to $v$ no event
from $\Gamma^+([v])$ was resampled. But this implies that $[v]$ was already
violated at parallel step $k$ and we did not select a maximal independent
set of violated events there for resampling. The contradiction shows that the depth of $\tau_k$
is indeed one more than that of $\tau_{k+1}$. To finish the proof notice
that $\tau_C(t)=\tau_1$. \qed

Let $Q(k)$ denote the probability, that the parallel algorithm makes at least
$k$ steps. By Lemma~\ref{para} some witness tree of depth $k-1$ must occur
in the log in this case. Notice that a depth $k-1$ witness tree has at
least $k$ vertices. Let $\mathcal{T}_A(k)$ be the set of witness trees in
$\mathcal{T}_A$ having at least $k$ vertices. We have
$$Q(k)\le\sum_{A\in\mathcal A}\sum_{\tau\in\mathcal{T}_A(k)}
\pr[\tau \mbox{ appears in the log } C]
\le \sum_{A\in \mathcal A}\sum_{\tau \in \mathcal{T}_A(k)}\prod_{v \in V(\tau)}
\pr[[v]]\le
(1-\varepsilon)^k\sum_{A\in\mathcal A}\sum_{\tau\in
\mathcal{T}_A}\prod_{v\in V(\tau)}x'([v]),$$
where the last inequality follows from the assumption in
Theorem~\ref{parallel}. Then, as before, we have
$$Q(k)\le(1-\varepsilon)^k\sum_{A\in\mathcal A}\sum_{\tau\in
  \mathcal{T}_A}\prod_{v\in
  V(\tau)}x'([v])=(1-\varepsilon)^k\sum_{A\in\mathcal
  A}\frac{x(A)}{1-x(A)}\sum_{\tau\in\mathcal{T}_A(k)}p_\tau\le
(1-\varepsilon)^k\sum_{A\in\mathcal A}\frac{x(A)}{1-x(A)}.$$
This bound easily implies Theorem~\ref{parallel}. \qed

\section{A deterministic variant}

It is possible to derandomize our algorithm under the additional
assumptions listed in Theorem~\ref{deterministic}. The idea is to create 
a list of all large potential witness trees which, if done with care, 
remains polynomial in length, and then to search for a table of evaluations 
of all random variables which, substituted for the random source in the original 
randomized variant, guarantees that the running time will be low.

To begin with, let us note that we can assume the weights $x(A)$ for
$A \in \mathcal{A}$ to be bounded away from $1$. If they are not, we can
simply replace $x(A)$ by, say, $\tilde x(A) := (1-\epsilon/2)x(A)$ for all
$A$ and $\epsilon$ by $\epsilon/2$. It is easily checked that all 
requirements are still satisfied and now the weights $\tilde x(A)$ are
upper bounded by $(1-\epsilon/2)$.

Let $(v_i^{(j)} \in D_i)_{1 \le i \le n, j \in \mathbb{N}}$ be sequences of values for the 
variables. Suppose we replace the random source by these sequences in such a way that
we take $v_i^{(j)}$ as the $j$-th sample $P_i^{(j)}$ and now run either the parallel or
the sequential algorithm as detailed in the previous sections. Recalling the proof of
Lemma~\ref{coupling}, let us say that a witness tree $\tau$ is \emph{consistent} with
the sequences $(v_i^{(j)} \in D_i)_{1 \le i \le n, j \in \mathbb{N}}$ if the $\tau$-check
passes when substituting them for the random source. We have proved that if the values
$v_i^{(j)}$ are selected at random, then the expected number of witness trees which are
consistent with them and which feature at least $k$ vertices is bounded by 
$(1-\varepsilon)^k\sum_{A\in\mathcal A}\frac{x(A)}{1-x(A)}$
and if the weights are bounded away from $1$, this is in turn bounded by
$\mathcal{O}(m(1-\varepsilon)^k)$. There exists therefore a constant $c$ such that
the expected number of witness trees of size at least $c \log m$ that are 
consistent with the sequences is at most $1/2$. With probability at least $1/2$, 
no consistent witness tree of size at least $c \log m$ exists at all.
Since no variable can then be reassigned more than $c \log m$ times, we can restrict ourselves to
finding values $(v_i^{(j)} \in D_i)_{1 \le i \le n, 0 \le j \le c \log m}$ such
that no large tree becomes consistent. Such values can be found sequentially
using the method of conditional expectations. First of all, we have to make sure
that the number of witness trees we have to consider is not too large, which is
demonstrated by the following lemma.

\begin{lemma}
\label{smallerwitnesses}
Suppose the maximum degree of the dependency graph is bounded by $k$ as required in the
theorem. Let $u \in \mathbb{N}$. If there exists a witness tree of size at least $u$ 
that is consistent with given evaluation sequences $(v_i^{(j)})_{1 \le i \le n, j \in \mathbb{N}}$, then there also exists
a witness tree of some size in the range $[u, (k+1)u]$ that is also consistent with the sequences.
\end{lemma}

\proof To arrive at a contradiction, assume that the claim is wrong for some value $u$. Then let
$\tau$ be a smallest counterexample, that is a consistent witness tree of size larger than $(k+1)u$ such
that there exists no consistent witness tree of a size in the range $[u,(k+1)u]$. Due to the bounded
maximum degrees of the dependency graph, each node in $\tau$ has at most $k+1$ children. Let $w_1, w_2, \mathellipsis, w_j$
with $j \le k+1$ be the immediate children of the root. Now we can build $j$ distinct witness trees
that are all consistent with the sequences as follows: traverse $\tau$ in the usual level-by-level fashion
bottom-up, starting at the lowest level and ending at the root. Consider this sequence of corrections
to be a log $C'$ of length $|\tau|$ and now construct for the $j$ penultimate correction steps in 
$C'$ (which obviously correspond to $w_1$ through $w_j$) the corresponding witness trees $\tau_i := \tau_{C'}(|\tau|-i)$
for $i=1,2, \mathellipsis, j$. For obvious reasons, each $\tau_i$ is consistent with the sequences.
Moreover, since $\tau_i$ must contain at least as many vertices as the subtree of $\tau$ rooted at $w_i$,
there must be at least one $i$ such that $\tau_i$ has at least $(|\tau|-1)/(k+1)$ vertices. Since
$\tau$ has at least $(k+1)u + 1$ vertices, $\tau_i$ has at least $u$ of them, implying that $\tau_i$ either
contradicts the assumption or constitutes a smaller counterexample. \qed

A deterministic algorithm can now proceed as follows. Enumerate all witness trees that have a size
in the range $[c \log m, (k+1) c \log m]$ in a list $L$. There is clearly a polynomial number of those given that
the dependency graph has bounded degrees. Now, in any ordering, go through all index pairs $(i,j)$
with $1 \le i \le n, 0 \le j \le c \log m$, computing suitable values for $v_i^{(j)}$ incrementally.
For each of them, consider each possible value of $D_i$ that we could assign to $v_i^{(j)}$ and then
compute the expected number of trees in $L$ that become consistent with the sequences of values given
that all values chosen so far and $v_i^{(j)}$ are fixed and the yet uncomputed values are considered
to be random. In the beginning when no value is fixed so far, that expected value is at most $1/2$.
Clearly, in each step we can preserve that it remains at most $1/2$ by selecting a value for $v_i^{(j)}$
that minimizes that expectation. Once all values are fixed, the expectation has to coincide with the
actual number of consistent witness trees in $L$ and therefore that number has to be zero.

Computing the probability of each tree $\tau \in L$ given that some
evaluations are random and others are fixed can be done easily by traversing $\tau$ in the usual
bottom-up level-by-level fashion (note that each time we encounter a variable during this traversal
we increment a counter pointing to the sample that has to be used as the current value for this
variable), for each vertex computing the conditional probability that the
corresponding event is violated given the values for the samples fixed so far (using the algorithm
we assume to exist in the hypothesis) and multiplying those probabilities.

After this polynomial preprocessing step, we run the usual algorithm substituting our tailored
values for the random source having guaranteed that it will terminate after a polynomial number
of steps. \qed

\section{The Lopsided Local Lemma}

In this section we study a lopsided version of the Local Lemma that is
slightly outside the framework of Theorem~\ref{lovasz}. Using lopsided
dependence, which is an adaptation of the notion in \cite{ES91} to our
setting, we formulate our main result in a slightly more general form. We
started with the original formulation of Theorem~\ref{thmmain}
because we find it more intuitive and it is general enough for most
applications of the Local Lemma. This lopsided generalization can also be
applied to the derandomized variant of Theorem~\ref{deterministic}. On the
other hand, we could not find an effective parallelization. 

We are still in the setting of Theorem~\ref{thmmain}: $\mathcal{P}$ is a finite set
of mutually independent random variables in a probability space and $\mathcal{A}$
is a finite set of events determined by these variables. We say that two
events $A,B\in\mathcal{A}$ are {\em lopsidependent} if there exist two evaluations $f$
and $g$ of the variables in $\mathcal{P}$ that differ only on variables in
$\mbox{vbl}(A)\cap\mbox{vbl}(B)$ such that $f$ violates $A$ and $g$ violates
$B$ but either $f$ does not violate $B$ or $g$ does not violate $A$. The {\em
lopsidependency graph} is the graph on the vertex set $\mathcal{A}$, where
lopsidependent events are connected by an edge. We write
$\Gamma'(A)=\Gamma'_{\mathcal{A}}(A)$ for the neighborhood of an event $A$ in
this graph.

Clearly, if $\mbox{vbl}(A)$ is disjoint from
$\mbox{vbl}(B)$, then $A$ and $B$ cannot be lopsidependent, so we have
$\Gamma'(A)\subseteq\Gamma(A)$. Substituting $\Gamma'(A)$ for $\Gamma(A)$ in
the statement of Theorem~\ref{thmmain} makes the assumption weaker and
therefore the theorem itself stronger.

We call an event $A\in\mathcal{A}$ {\em elementary}, if there is a single
evaluation of the variables in $\mbox{vbl}(A)$ violating $A$. Lopsidependence
between elementary events is more intuitive: the elementary events $A$ and
$B$ are lopsidependent if they are disjoint (mutually exclusive). Elementary
events form a very special case of events considered in the Local Lemma, but
in case the variables in $\mbox{vbl}(A)$ have finite domain, any event $A$
is the union of finitely many elementary events determined by the same set of
variables. Avoiding $A$ is equivalent to avoiding all of these elementary
events and such a ``breaking up'' of event $A$ is not too costly in the
following sense. If the assignment $x:\mathcal{A}\to(0,1)$ satisfies the
(slightly stronger than usual) condition
$$ \forall A \in \mathcal{A} \; : \; \pr[A] \; \le \; x(A) \prod_{B \in
\Gamma^+(A)} (1 - x(B)),$$
then we can break up all the events $A\in\mathcal{A}$ into elementary events and
find a corresponding assignment to the elementary events satisfying the same
condition. Even in cases where a suitable assignment is not possible for the set
$\mathcal{A}$, breaking up the events may cause many of the dependencies
among the elementary events to disappear when considering lopsidependence, and
Theorem~\ref{lopsiversion} may become applicable.

An important application where using lopsidependence has proven to be effective is
the satisfiability problem. In a CNF formula, two clauses $A$ and $B$ are said to have a
\emph{conflict} if one contains a literal $l$ and the other one its complement $\bar l$.
When conducting the random experiment of sampling a truth assignment to each variable
independently, the (elementary) events that $A$ is violated or that $B$ is violated,
respectively, are lopsidependent on one another. If, on the other hand, $A$ and $B$
merely \emph{overlap} in some literal, then correcting either of them in the case of
being violated will not harm the other one. Hence, being in conflict is the more relevant
notion to analyze dependencies in a SAT problem. The lopsided Local Lemma has been
effectively used, e.g. by Berman, Karpinski and Scott in \cite{BKS03}, to prove better
bounds on the number of occurrences per variable that can be allowed while guaranteeing 
a satisfying assignment. Proofs of this type can be made constructive using the method we 
present here.

\begin{theorem}
\label{lopsiversion}
Let $\mathcal P$ be a finite set of mutually independent random variables in a
probability space. Let $\mathcal{A}$ be a finite set of events determined by
these variables. If there exists an assignment of reals $x : \mathcal{A} \rightarrow (0,1)$ such that
$$ \forall A \in \mathcal{A} \; : \; \pr[A] \; \le \; x(A)
\hspace*{-0.2cm} \prod_{B \in \Gamma'_{\mathcal A}(A)} (1 - x(B)),$$
then there exists an assignment of values to the variables $\mathcal P$
not violating any of the events in $\mathcal A$. Moreover our
randomized algorithm resamples an event $A \in \mathcal{A}$ at
most an expected $x(A)/(1-x(A))$ times before it finds such an
evaluation. Thus the expected total number of resampling steps is at most
$\sum_{A \in \mathcal{A}} \frac{x(A)}{1-x(A)}.$ 
\end{theorem}

As the proof of this theorem is almost identical to that of
Theorem~\ref{thmmain} we concentrate on the few differences.
We define
$\Gamma'^+(A)=\Gamma'^+_{\mathcal{A}}(A)=\Gamma'_{\mathcal{A}}(A)\cup\{A\}$ for
$A\in\mathcal{A}$.
Given a log $C$ of an execution of Algorithm~1.1 and a time index $t$, we
define the \emph{lopsided witness tree} $\varrho_C(t)$ similarly to the
usual witness tree $\tau_C(t)$, but now the children of a
vertex labelled $A$ will be labelled from $\Gamma'^+(A)$. Formally we start
with the root labeled $C(t)$ and going through the time steps $i=t-1, t-2,
\mathellipsis, 1$ of the log we add a new vertex labelled $C(i)$ every time
$C(i)\in\Gamma'^+(A)$ for a label $A$ of a vertex of the tree constructed so
far. If such a vertex exists we
choose one as far from the root as possible and
attach the new vertex of label $C(i)$ to this existing vertex.

We denote the label of a vertex $v$ by $[v]$ and call a witness tree a {\em
proper lopsided witness tree} if the children of any vertex $v$ receive
distinct labels from $\Gamma'^+([v])$.
Clearly, $\varrho_C(t)$ is a proper lopsided witness tree, but the equivalent
of part (ii) of Lemma~\ref{coupling} is less trivial. We state and prove it
below. One also has to modify the Galton-Watson branching
process considered to produce a proper lopsided witness tree. With these
modifications the original proof of Theorem~\ref{thmmain} carries over to the
lopsided version.

\begin{lemma}\label{loplemma}
Let $\tau$ be a fixed proper lopsided witness tree and $C$ the (random) log
produced by the algorithm. The probability that $\tau=\varrho_C(t)$ holds for
some $t$ is at most $\prod_{v\in V(\tau)}\pr[[v]]$.
\end{lemma}

\proof We prove this statement with the same coupling argument as in
Lemma~\ref{coupling}: if $\tau=\varrho_C(t)$, then the $\tau$-check (as
defined in the proof of Lemma~\ref{coupling})
passes if both the algorithm and the $\tau$-check use the same list of
independent random samples $P^{(0)},P^{(1)},\ldots$ for the variables
$P\in\cal P$. Recall that the $\tau$-check considers the vertices of $\tau$ in
some fixed order $v_1,\ldots,v_s$ of non-increasing distance from the root
and, when considering $v_i$ draws new samples for the variables in
$\mbox{vbl}([v_i])$ and checks if they violate $[v_i]$. Clearly, the
probability that the $\tau$-check passes (meaning that each event is violated
when checked) is exactly $\prod_{v\in V(\tau)}\pr[[v]]$.

Let us fix the
sequences $P^{(0)}, P^{(1)},\ldots$ of random samples of the
variables $P\in\mathcal{P}$. The log of our algorithm is still not determined
because of the arbitrary choices we make in selecting the violated event to
resample. If a possible log $C$ satisfies $\varrho_C(t)=\tau$ for some $t$ we denote by
$q_C(v)$ the time step the vertex $v$ was added during the construction
of $\varrho_C(t)$. In particular we have $q_C(r)=t$ for the root $r$ of $\tau$
and we have $C(q_C(v))=[v]$.
If there are several possible logs $C$ with $\tau=\varrho_C(t)$ for some $t$,
then we choose the one that minimizes
$w(C):=\sum_{i=1}^s(s+1-i)q_C(v_i)$. We claim that with this choice we have
$q_C(v_i)=i$ for $i=1,\ldots,s$. Assuming the claim it is clear that the
$\tau$-check passes since when it considers $v_i$ it will draw the same random
values for the variables in $\mbox{vbl}([v_i])$ that makes the selection of
$C(i)=[v_i]$ a valid choice for resampling in the $i$th step of our algorithm.

To prove the claim above notice first that the definition of lopsidependence
ensures that if $C$ is a possible log of our algorithm and for some $j$
the events $C(j)$ and
$C(j+1)$ are not lopsidependent, then the log $C'$ obtained
from $C$ by reversing the order of these two resamplings is still
possible. Furthermore if $\tau$ can be obtained as $\varrho_C(t)$ for some $t$
then $\tau$ can also be obtained as $\varrho_{C'}(t)$ for some $t$. Now assume
that the claim does not hold. In this case there must exist indices $i$ and
$j$ with $q_C(v_i)=j+1$ and $q_C(v_{i'})\ne j$ for any $i'<i$. But this means that
during the construction of $\tau=\varrho_C(t)$ when considering time step $j$
we did not attach a new vertex labeled $C(j)$, or if we did, it went on the
level of $v_i$ or even higher. In either case $C(j)$ must not be
lopsidependent from $C(j+1)=[v_i]$, neither can $C(j)$ and $C(j+1)$
coincide. Thus the log $C'$ obtained from $C$ by switching $C(j)$ and $C(j+1)$
is also a possible log producing $\tau$ as a lopsided witness tree, and we
also have $w(C')<w(C)$, a contradiction. \qed

\section{Conclusion}

The bound $x(A)/(1-x(A))$ in Theorem~\ref{thmmain} is tight, but it is only
achieved in the rather uninteresting case, when $A$ is an isolated vertex of
the dependency graph and we set $x(A)=\pr[A]$.
Consequently the bound on the total number of
resampling steps is achieved only if all the events in $\mathcal A$ are
independent. To see that the expected number of resamplings of an event
$A \in \mathcal A$ cannot achieve the bound $x(A)/(1-x(A))$ unless $A$ is isolated
in the dependency graph notice that for equality any proper witness tree with
the root labelled $A$ must be present in the log with positive probability,
but only those can be present where the labels on each level form an
independent set in the dependency graph.

To implement the sequential version of our algorithm we need only to assume
that we have an algorithm sampling the variables in $\mathcal{P}$
and another that finds the set of violated events for a given evaluation
of the variables. For the parallel version we also need to be able to find a
maximal independent set of violated vertices of the dependence graph. Luby's
randomized algorithm
finds a maximal independent set in any graph in logarithmic expected time
using a processor associated with each vertex of the graph. As we have to find
an expected logarithmic number of maximal independent sets, one after the
other, the total running time of this part of the algorithm is
$O(\log^2 m)$, where $m=|V(G)|=|\mathcal A|$. Sampling and finding the
violated indices is typically faster, so this might be the bottleneck of our
algorithm. One can reduce the time required to find an independent set to a
constant by implementing only a single step of Luby's algorithm. The result
will not be a maximal independent set, nevertheless we can use it to perform
the corresponding resamplings in parallel, but the expected number of steps
this version of the algorithm takes is yet to be analyzed.

Finally, it remains an interesting open question whether it is possible to
derandomize our algorithm even if the degrees of the dependency graph are
unbounded.

\section*{Acknowledgements}

Many thanks go to Dominik Scheder and Philipp Zumstein for various very helpful comments and to Emo Welzl for many fruitful discussions 
and the continuous support.

\end{document}